\documentclass[12pt,draftclsnofoot, perreview, onecolumn]{IEEEtran}
\usepackage{graphicx}
\usepackage{amsmath,amssymb}
\usepackage{mathrsfs}
\usepackage{cases}
\usepackage{color}
\usepackage{amsfonts}
\usepackage{indentfirst}
\usepackage{cite}
\usepackage{caption}
\usepackage{algorithm}
\usepackage{algpseudocode}
\usepackage{booktabs}
\usepackage{subfigure}
\usepackage{multirow}
\usepackage{amsthm}
\usepackage{diagbox}
\newtheoremstyle{mythm}{3pt}{3pt}{}{16pt}{\bfseries}{:}{.5em}{}
\theoremstyle{mythm}

\newcommand{\tabincell}[2]{\begin{tabular}{@{}#1@{}}#2\end{tabular}}

\usepackage{threeparttable}

\begin{document}
\title{Design of Coded Caching Schemes with Linear
Subpacketizations Based on Injective Arc Coloring of Regular Digraphs}

\author{Xianzhang Wu, Minquan Cheng, Li Chen, {\em Senior Member, IEEE}, Congduan Li, {\em Member, IEEE}, and Zifan Shi
\thanks{Xianzhang Wu and Congduan Li are with the School of Electronics and Communication Engineering, Sun Yat-sen University, Shenzhen 518107, China (e-mail: wuxzh7@mail2.sysu.edu.cn, licongd@mail.sysu.edu.cn).}
\thanks{Minquan Cheng and Zifan Shi are with the Guangxi Key Lab of Multi-source Information Mining $\&$ Security, Guangxi Normal University,
Guilin 541004, China (e-mail: chengqinshi@hotmail.com, tianle19951116@hotmail.com).}
\thanks{Li Chen is with the School of Electronics and Information Technology, Sun Yat-sen University, Guangzhou 510006, China (e-mail: chenli55@mail.sysu.edu.cn).}
}


\maketitle

\begin{abstract}
Coded caching is an effective technique to decongest the amount of traffic in the backhaul
link. In such a scheme, each file hosted in the server is divided into a number of packets to pursue a low transmission rate based on the delicate design of contents cached into users and broadcast messages. However, the implementation complexity of this scheme increases with the number of packets. It is desirable to design a scheme with a small subpacketization level and a relatively low transmission rate. Recently, placement delivery array (PDA) was proposed to address the subpacketization bottleneck of coded caching. This paper investigates the design PDA from a new perspective, i.e., the injective arc coloring of regular digraphs. It is shown that the injective arc coloring of a regular digraph can yield a PDA with the same number of rows and columns. Based on this, a new class of regular digraphs are defined and the upper bounds on the injective chromatic index of such digraphs are derived. Consequently, some new coded caching schemes with a linear subpacketization level and a small transmission rate are proposed, one of which generalizes the existing scheme for the scenario with a more flexible number of users.
 \end{abstract}


\IEEEpeerreviewmaketitle
\section{Introduction}\label{introduction}
\IEEEPARstart{T}{he}
  dramatic increase of video streaming requests can easily cause severe network congestions during the peak-traffic times. One possible solution is to exploit the off-peak network resources, such as to cache some of the possibly demanded contents in users' local memories, i.e., the so called caches. This is a natural way to utilize each user's own cache to decrease the network traffic when the cached contents are requested. The gain offered by this approach is called {\it local gain}, which depends on the size of local caches. A more effective way of caching is through coding, which was first proposed by Maddah-Ali and Niesen (MN) \cite{1}. It reduces the network pressure during the peak times by strategically designing the contents cached into network users and the broadcast messages to obtain the {\it global gain}. In the centralized coded caching system,
a central server containing $N$ files of the same size is connected to $K$ users over a noiseless shared link. Each user has a cache memory with a size of $M$ files, where $M<N$. It operates in two phases: the placement phase during the off-peak times and the delivery phase during the peak times. In the placement phase, each file is divided into $F$ equal packets, and each user's cache is filled with some form of these packets without any prior knowledge
of future demands. The quantity $F$ is referred to as the {\it subpacketization level}. In the delivery phase, each user reveals its requested
file to the server. After receiving the user demands, the server transmits some coded symbols over a noiseless shared link to all the users so that their demands can be satisfied with the assistance of the locally cached contents.
Normalizing the minimal worst case transmission load by the size of file leads to the so called {\it transmission rate} $R$, i.e., the minimum number of files that must be communicated so that any possible demands can be satisfied.
Under such paradigm, if the packets are cached directly without coding in the placement phase, it is called an {\it uncoded placement}; otherwise, it is called a {\it coded placement}.

The MN scheme is realized by a combinatorial design in the placement phase and a linear
coding in the delivery phase such that each multicast message satisfies the demands of multiple users. It achieves an optimal transmission rate under the constraints of uncoded placement and $K\leq N$ \cite{2}. Observing that there exist some redundant transmissions in the MN scheme when a file is requested by several users, Yu {\it et al.} \cite{0400} derived a scheme that improves the MN scheme and achieves the optimal transmission rate. They further showed that the multiplicative gap between the optimal caching scheme with uncoded placement and any caching scheme with coded placement is at most two \cite{3}.
The MN scheme has been extensively studied over other network scenarios, such as the decentralized caching \cite{003}, the multi-level popularity and access \cite{558}, the combination networks \cite{559} and the device-to-device (D2D) caching systems \cite{557}.
However, the subpacketization level of the MN scheme increases exponentially with the number of users, which makes it impractical for large networks. It is important to reduce the subpacketization level of the MN scheme, while maintaining a relatively low transmission rate.
\begin{table}[H]
\center
\caption{Summary of Some Known Coded Caching Schemes\label{tab-known-1}}
\begin{threeparttable}
\renewcommand\arraystretch{1}
  \setlength{\tabcolsep}{0.40mm}{
\begin{tabular}{|c|c|c|c|c|c|}
\hline
Schemes and Parameters & User Number $K$  & Caching Ratio $\frac{M}{N}$
& Rate $R$   & Subpacketization Level $F$    \\ \hline

\tabincell{l}{MN scheme in \cite{1},  any $k$, $t\in  \mathbb{N}^+$ \\with $t<k$} &$k$&$\frac{t}{k}$&$\frac{k-t}{1+t}$&${k\choose t}$ \\ \hline

\tabincell{c}{Scheme in \cite{14}, any $a,b,m,\lambda\in\mathbb{N}^{+}$\\ with $a<m$,$b<m$ and $\lambda < \min{\{a,b\}}$}& ${m \choose a}$& $1-\frac{{a \choose \lambda}{m-a \choose b-\lambda}}{{m \choose b}}$& $\frac{{m \choose a+b-2\lambda}{a+b-2\lambda \choose a-\lambda}}{{m \choose b}}$
&${m \choose b}$\\ \hline

\tabincell{c}{Scheme in \cite{22}, any $n,m,k\in\mathbb{N}^{+}$,\\ prime power $q$ with $n+m\leq k$}
& $\frac{q^{\frac{n(n-1)}{2}}\prod_{i=0}^{n-1}\left[k-i \atop 1\right]_q}{n!}$
&\tabincell{c}{ $1-q^{mn}\cdot$\\ $\prod_{i=0}^{n-1}\frac{\left[k-m-i \atop 1\right]_q}{\left[k-i \atop 1\right]_q}$}
&\tabincell{c}{ $\frac{m!q^{mn}}{(m+n)!}q^{\frac{n(n-1)}{2}}\cdot$\\ $\prod_{i=0}^{n-1}\left[k-m-i \atop 1\right]_q$}
& $\frac{q^{\frac{m(m-1)}{2}}\prod_{i=0}^{m-1}\left[k-i \atop 1\right]_q}{m!}$
\\ \hline


\tabincell{c}{Scheme in \cite{50}, any $n,w'\in\mathbb{N}^{+}$ \\with $w'<n$}&$2^{n}$ &$1-\frac{{n\choose w'}}{\sum_{i=0}^{w'} {n\choose i}}$&$ \frac{{{n\choose w'} 2^{n-w'}}}{\sum_{i=0}^{w'} {n\choose i}}$
&$\sum_{i=0}^{w'} {n\choose i}$\\  \hline


\tabincell{c}{Scheme in \cite{17}, $\epsilon(\delta)\rightarrow 0$ and\\ $k(\delta)\rightarrow \infty$ as $\delta\rightarrow0$} &$k(\delta)$ &$k^{-\epsilon(\delta)}$ &$k^{\delta}$& $k(\delta)$\\\hline

\tabincell{c}{Scheme in \cite{80000}, any $r,k,z\in\mathbb{N}^{+}$} &$2^{r}k$ &$1-\frac{r+1}{2^{r}}+\frac{rz}{2^{r}k}$ &$\frac{k^{2}+k^{2}r-krz}{2^{r}k}$& $2^{r}k$\\\hline
\end{tabular}}
\begin{tablenotes}
\item[*]In Table I, $\left[\alpha \atop \beta\right]_q=\frac{(q^{\alpha}-1)\cdots (q^{\alpha-\beta+1}-1)}{(q^{\beta}-1)\cdots (q-1)}$ for any positive integers $\alpha, \beta$ and prime power $q$.
\end{tablenotes}
\end{threeparttable}
\end{table}

 There exist some works on reducing the subpacketization level of the MN scheme, but they trade it with the transmission rate
 \cite{20,22,21,16,17,26,13,14,39,19,29,24,23,50,33,560,80000,500}. In particular, Yan {\em et al}. \cite{13} represented the coded caching scheme by an array called the placement delivery array (PDA) to reduce the subpacketization level. It has been shown that the MN scheme can be considered as a special class of PDAs. By the construction of PDAs, two new classes of coded caching schemes were obtained with a reduced subpacketization level over that of the MN scheme. But they yield a slightly increased transmission rate. Since then, PDA has been utilized as a systematic approach to design coded caching schemes that yield a low subpacketization level \cite{14,39,19,29,24,23,33,50,560,80000}.
Apart from PDA, other coded caching schemes with a reduced subpacketization level were designed
 by the use of Ruzsa-Szemer\'{e}di graphs \cite{17}, projective geometry and line graphs \cite{22}, hypergraphs \cite{16} and combinatorial design  \cite{20}.
 Table I summarizes the existing schemes with the advantages in either the subpacketization level or the transmission rate.

One of the most interesting areas for the coded caching problem is the scenario with a linear subpacketization level, i.e., the subpacketization increases linearly with the number of users.
It has been shown that with a small memory ratio requirement, a coded caching scheme with a linear subpacketization level can be constructed at a near constant transmission rate \cite{17}. However, it requires an extremely large number of users.
The MN scheme itself achieves $F=K$ when $\frac{M}{N}=\frac{1}{K}$, but the transmission rate is $R=\frac{K-1}{2}$.
The schemes proposed in \cite{50} can yield a linear subpacketization level and a small transmission rate. However, they require the number of users to be some non-flexible values. A more recent work was proposed in \cite{22}, where the scheme achieves a linear subpacketization level but with a larger transmission rate and a non-flexible number of users.
 Therefore, most coded caching schemes that yield a linear subpacketization level are disadvantageous in either the number of users or the transmission rate.

 This paper considers the PDA construction from the perspective of graph coloring, aiming
to design a coded caching scheme that can work for a flexible number of users with a linear subpacketization level and a small transmission rate.
Our technical contributions include:

$\bullet$ We integrate the PDA in coded caching with injective arc coloring of regular digraphs. It is shown that the injective arc coloring of a regular digraph can yield a PDA with the same number of rows and columns. This enables the design of coded caching schemes utilizing the existing structures of regular digraphs. By observing that the strong edge coloring of regular graphs can be viewed as a special injective arc coloring of regular digraphs, a new coded caching scheme  (as stated in {\it Theorem 4}) with a linear subpacketization level can be obtained from the existing strong edge coloring of unitary Cayley graphs.

$\bullet$ We also define a new class of regular digraphs and derive upper bounds on the injective chromatic index of such digraphs. Consequently, a new coded caching scheme (as stated in {\it Theorem 5}) with a flexible number of users are obtained, which yield a linear subpacketization level and a small transmission rate. Based on the proposed coded caching scheme, this research finds out that
some packets cached by the users have no multicast opportunities in the delivery phase. By utilizing the maximum distance separable (MDS) code in the placement phase, a new coded caching scheme (as stated in {\it Theorem 6}) with a smaller subpacketization level and memory ratio is further proposed, which generalizes the scheme of \cite{50} and supports a more flexible number of users

The rest of this paper is organized as follows. In Section II, we briefly review the background of the centralized coded caching system. The relationship between PDA and injective arc coloring of regular digraphs is presented in Section III. Section IV defines a new class of regular digraphs and presents upper bounds on the injective chromatic index of such digraphs. The new PDA schemes are proposed in Section V. Their performance analyses are given in Section VI. Finally, Section VII concludes the paper.
\section{System Model and the PDA}
This section presents the coded caching system model and the PDA. Some key notations are introduced as follows.

\textbf{Notations:} Let calligraphic symbols, bolded capital letters and bolded lower-case letters denote sets, arrays and vectors, respectively. Symbol $\oplus$ represents the exclusive-or (XOR) operation. We use $| \cdot |$ to denote the cardinality of a set. Let $\mathbb{N}^{+}$ denote the set of positive integers. The sets of consecutive integers are denoted as $[x: y]=\{x, x+1,\ldots, y\}$. Given an array \textbf{P}, let $\textbf{P}(i,j)$ denote its entry of row $i$ and column $j$.
 Let ${[0: m-1]\choose t}$ denote the collection of all subsets of $[0: m-1]$ with size $t$, i.e., ${[0: m-1]\choose t}=\{\mathcal{S}\mid\mathcal{S}\subseteq[0: m-1], |\mathcal{S}|=t\}$. For a length-$m$ vector $\textbf{x}$ and a set $\mathcal{S}\subseteq[0: m-1]$, let $\textbf{x}|_{\mathcal{S}}$ denote a vector obtained by taking the coordinate indexed by $j\in\mathcal{S}$.
\subsection{Centralized Coded Caching System}
In a centralized coded caching system, a server
containing $N$ files of the same size is connected to $K$ users through a noiseless shared link, as shown in Fig.1. Each
user is equipped with a
dedicated cache with a size of $M$ files, where $M < N$. The $N$ files and $K$ users are denoted by
$\mathcal{W} =\{W_0, W_1, \ldots, W_{N-1}\}$ and $\mathcal{K} =[0: K-1]$, respectively. An $F$-division $(K, M, N)$ coded caching scheme consists of two phases, which are described as follows.

$\bullet$ \textbf{Placement Phase}: Each file is divided into $F$ equal packets
 i.e., $W_n = \{W_{n,j}| j \in[0: F-1]\}$, $n\in[0:N-1]$. The
server places some packets (or coded packets) directly into each user's cache without any prior knowledge of the
demands in the delivery phase. Let $\mathcal{Z}_k$ denote the contents cached by user $k$, where $k\in \mathcal{K}$. The size of $\mathcal{Z}_k$ cannot be greater than the capacity of each user's cache memory size $M$.
\begin{figure}[H]
\begin{center}
\includegraphics[scale=0.23]{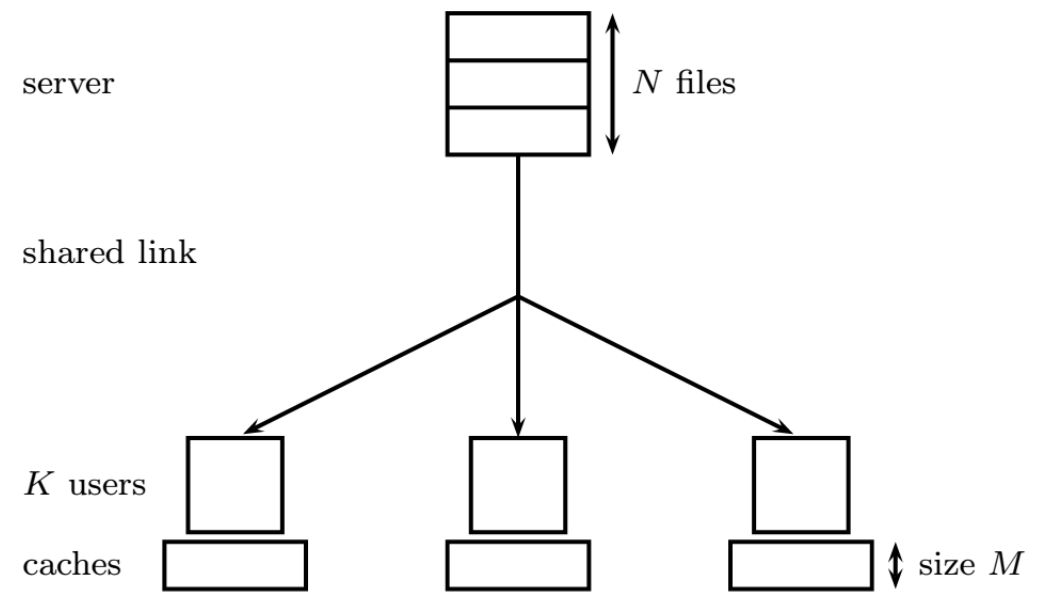}
\caption{\label{2}Coded caching system.}
 \end{center}
 \end{figure}

$\bullet$ \textbf{Delivery Phase}: Each user requests an arbitrary file from $\mathcal{W}$. The request vector is denoted by
$\textbf{d}=(d_0, d_1, \ldots, d_{K-1})$, i.e., user $k$ requests file $W_{d_k}$, where $k\in \mathcal{K}$ and $d_k\in[0: N-1]$. Once the server receives the request vector $\textbf{d}$, it broadcasts a signal of at most $RF$ packets such that all the users can correctly decode
their requested file together with the cached contents.
\subsection{Placement Delivery Array}
 Let us review the definition of PDA that can be used to characterize both the placement phase and the delivery
phase.

\emph{\textbf{Definition 1}} \cite{13}. Given $K, F, Z, S\in\mathbb{N}^{+}$, an $F\times K$ array $\textbf{P}=(\textbf{P}(i,j))$, where $i\in[0: F-1]$, $j\in[0: K-1]$, and $\textbf{P}(i,j)\in[0:S-1]\cup\{\ast\}$, is called a $(K, F, Z, S)$ PDA if it satisfies the following conditions:

C1. Symbol $``\ast"$ appears exactly $Z$ times in each column;

C2. Each integer of $[0:S-1]$ appears at least once in the array;

C3. For any two distinct entries $\textbf{P}(i_1,j_1)$ and $\textbf{P}(i_2,j_2)$, $\textbf{P}(i_1,j_1) = \textbf{P}(i_2,j_2) = s$ is an integer only if

(a). $i_1\neq i_2, j_1 \neq j_2$, i.e., they lie in distinct rows and distinct columns;

(b). $\textbf{P}(i_1,j_2) = \textbf{P}(i_2,j_1) = \ast$, i.e., the corresponding $2\times2$ subarray formed by rows $i_1, i_2$ and columns $j_1, j_2$ must be in one of
the following forms
\begin{equation*}
\left(
  \begin{array}{cc}
    s & \ast \\
    \ast & s \\
  \end{array}
\right) \text{,}\; \left(
  \begin{array}{cc}
    \ast & s \\
    s & \ast \\
  \end{array}
\right).
\end{equation*}

%

Algorithm 1 was proposed to realize the PDA based coded caching schemes. Given a $(K, F, Z, $\\$S)$ PDA \textbf{P} with column indices representing the users and row indices representing
the packets, if $\textbf{P}(j,k)= \ast$, it implies that the server has placed the $j$th packet of all the files into the cache of user $k$. Condition C1 of {\it Definition 1} implies that each user has the same memory size and the memory ratio is $\frac{M}{N}=\frac{Z}{F}$. If $\textbf{P}(j,k)=s$, where $s\in [0:S-1]$, it
indicates that the user $k$ does not cache the $j$th packet of all the files.
 The linear combination of the requested packets indicated by $s$ will be broadcast by the server at time slot $s$. Condition C3 of {\it Definition 1} ensures the decodability, since it has cached all the other packets in the multicast message except its requested one. Finally,
Condition C2 of {\it Definition 1} implies that the number of messages transmitted by the server is exactly $S$ and the transmission rate is $R =\frac{S}{F}$.
Based on Algorithm 1, an $F$-division $(K, M, N)$ coded caching scheme can be characterized by the following lemma.

\begin{algorithm}[H]
\caption{Coded Caching Scheme Based on PDA \cite{13}}
\textbf{1:} \textbf{Procedure} Placement (\textbf{P}, $\mathcal{W}$)\\
\textbf{2:} $\;\;\;\;$Split each file $W_n\in \mathcal{W}$ into $F$ packets as $W_n=\{W_{n,j}\mid j\in[0: F-1]\}$.\\
\textbf{3:} $\;\;\;\;$\textbf{For} {$k\in\mathcal{K}$} \textbf{do}\\
\textbf{4:} $\;\;\;\;\;\;$ $\mathcal{Z}_k\leftarrow\{W_{n,j}\mid \textbf{P}(j,k)=\ast, \forall n\in[0: N-1]\}$;\\
\textbf{5:} \textbf{Procedure} Delivery$\;(\textbf{P}, \mathcal{W}, \textbf{d})$\\
\textbf{6:} $\;\;\;\;$\textbf{For} {$s=0, 1, \ldots, S-1$} \textbf{do}\\
\textbf{7:} $\;\;\;\;\;\;$ Server sends $\oplus_{{\textbf{P}(j, k)=s, j\in[0: F-1], k\in[0:K-1]}}W_{d_k, j}$.
\label{code:recentEnd}
\end{algorithm}



\emph{\textbf{Lemma 1}} \cite{13}. Given a $(K, F, Z, S)$ PDA, there always exists an $F$-division $(K, M, N)$ coded caching scheme with a memory ratio of $\frac{M}{N}=\frac{Z}{F}$ and a transmission rate of $R =\frac{S}{F}$.

The following example demonstrates this property.

\emph{\textbf{Example 1}}. Given a (4, 4, 2, 4) PDA \textbf{P}. Based on Algorithm 1, a 4-division (4, 2, 4) coded caching scheme can be obtained as

\begin{equation}
\textbf{P}=\left(
  \begin{array}{cccc}
    0 & \ast & \ast& 3\\
    \ast & 1& 2& \ast \\
    1 &\ast& \ast& 2 \\
   \ast &0& 3&\ast \\
  \end{array}
  \right).
\end{equation}


$\bullet$ \textbf{Placement Phase}: Each file $W_n$ is divided into four packets, i.e., $W_n=\{W_{n,0}, W_{n,1}, W_{n,2}, W_{n,3}\}$, where $n\in[0: 3]$. The contents cached by each user are
$$\mathcal{Z}_0=\{W_{n,1}, W_{n,3}\mid n\in[0: 3]\}; \mathcal{Z}_1=\{W_{n,0}, W_{n,2}\mid n\in[0: 3]\};$$
$$\mathcal{Z}_2=\{W_{n,0}, W_{n,2}\mid n\in[0: 3]\}; \mathcal{Z}_3=\{W_{n,1}, W_{n,3}\mid n\in[0: 3]\}.$$

$\bullet$ \textbf{Delivery Phase}: Let us assume that the request vector is $\textbf{d}=(0, 1, 2, 3)$. The signals sent by the server at the
four time slots (TSs) are listed as follows. TS-0: $W_{0,0}\oplus W_{1,3}$; TS-1:$W_{0,2}\oplus W_{1,1}$;
TS-2: $W_{2,1}\oplus W_{3,2}$; TS-3: $W_{2,3}\oplus W_{3,0}$. Each user can then reconstruct its required file. E.g., user $0$ requires $W_0$ and it has cached $W_{0,1}$ and $W_{0,3}$. At TS-0, it can obtain $W_{0,0}$ with its received coded packet $W_{0,0}\oplus W_{1,3}$, where $W_{1,3}$ was cached. At TS-1, it can obtain $W_{0,2}$ with its received coded packet $W_{0,2}\oplus W_{1,1}$, where $W_{1,1}$ was cached. Hence, the transmission rate is $R=\frac{4}{4}=1$.
\section{Injective Arc Coloring of Regular Digraphs and Its Relation to PDA}
This section investigates the design of a PDA with the same number of rows and columns through the injective arc coloring of a regular digraph.
 Some graph theoretic notations are reviewed as follows. Let $G=G(\mathcal{V}, \mathcal{E})$ denote a simple  undirected graph with vertex set $\mathcal{V}$ and edge set $\mathcal{E}$. The degree of a vertex $v$ in a graph $G$ is denoted by $d(v)$. A graph $G$ is called $r$-regular if $d(v) = r$ for all $v\in\mathcal{V}$. Let $D=D(\mathcal{V},\mathcal{E})$ denote a digraph with vertex set $\mathcal{V}$ and arc set $\mathcal{E}$. For a vertex $v\in\mathcal{V}$, we denote the indegree and outdegree of $v$ by $d^{-}(v)$ and $d^{+}(v)$, respectively. If $d^{-}(v)=d^{+}(v)=r$ for each vertex $v\in\mathcal{V}$, $D$ is called a $r$-regular digraph. In this paper, we focus on the regular digraph with reverse arcs but without a  directed self loop. For clarity, we introduce several definitions of graph coloring.

\emph{\textbf{Definition 2}}\cite{564}. For a graph $G$, a proper edge coloring is an assignment of colors to each edge of a graph such that no
two edges with a common endpoint receive the same color. The smallest number of colors needed in a proper edge coloring
of a graph $G$ is called the chromatic index of $G$.

\emph{\textbf{Definition 3}}\cite{564}. A strong edge coloring is a proper edge coloring, with the further condition that no two edges with the same color lie on a path of length three. The strong edge chromatic number is the minimum number of colors that allow a strong edge coloring, denoted by $\chi_s(G)$.

The concept of injective edge coloring was first introduced by Cardoso {\em et al}. \cite{563}. Its definition is described as follows.

\emph{\textbf{Definition 4}} \cite{563}. An edge coloring of a graph $G$ is injective if any two edges $e$ and $f$ that are at a distance of
exactly one (i.e., there exists an edge between $e$ and $f$) or in a common triangle receive distinct colors. The injective chromatic index of $G$, denoted by $\chi_{i}(G)$, is the minimum number of colors needed for an injective edge coloring of $G$.

It can be seen that a strong edge coloring of a graph $G$ is an injective edge coloring, but not vice versa.
In this paper, we extend the definition of injective edge coloring of graph $G$ to the regular digraph $D$ as follows.

\emph{\textbf{Definition 5}}. An arc coloring of a regular digraph $D$ is injective if any two arcs \textbf{e} and \textbf{f} that are at a distance of exactly one (i.e., there exists an arc between two arcs \textbf{e} and \textbf{f} in a directed path) or in a common directed triangle receive distinct colors. The injective chromatic index of $D$, denoted by $\chi_{i}(D)$, is the minimum number of colors needed for an injective arc coloring of $D$.

The following example illustrates the above definition.


\emph{\textbf{Example 2}}. Given the following arc-colored regular digraph, based on {\it Definition 5}, it can be observed that this coloring is an injective arc coloring.
\begin{figure}[H]
\begin{center}
\includegraphics[scale=0.56]{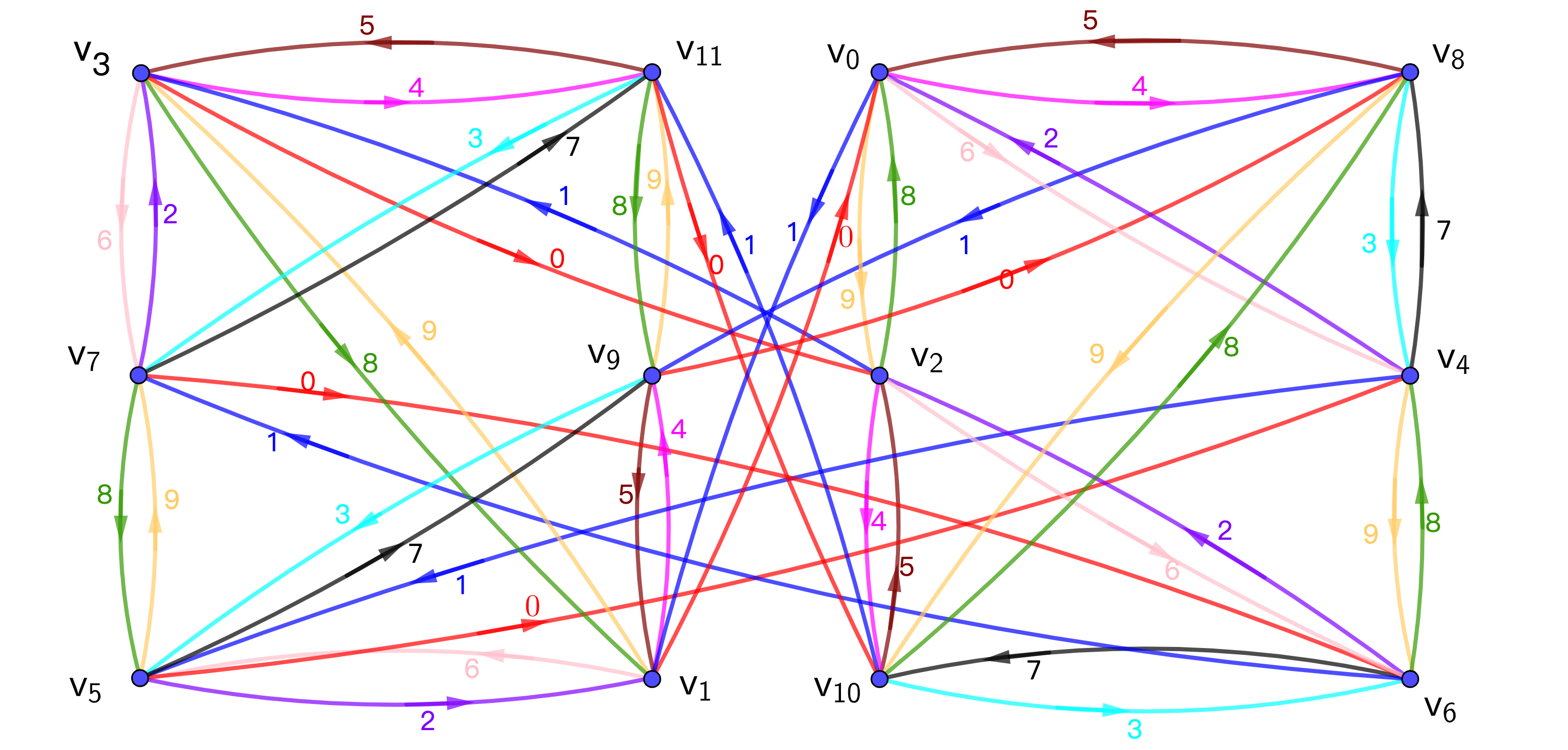}
\caption{\label{2} 4-regular digraph with the number of injective arc coloring of 10.}
 \end{center}
 \end{figure}

Based on the above introduction, given an injective arc-colored regular digraph
$D$ with vertex set $\mathcal{V}$ and arc set $\mathcal{E}$, if the arcs in $\mathcal{E}$ are colored by the colors $0, 1, \ldots, S-1$, we can construct a $|\mathcal{V}|\times |\mathcal{V}|$ array $\textbf{P}=(\textbf{P}(v_j,v_k))$ $(v_j,v_k\in\mathcal{V})$ composed of alphabet set $[0:S-1]\cup\{\ast\}$ as

\begin{equation}
        \textbf{P}(v_j,v_k)= \begin{cases}
         \ast, & \text{if $(v_j,v_k)\notin\mathcal{E}$};\\
         s, &\text{if $(v_j,v_k)\in\mathcal{E}$ and it is colored by $s$.}
 \end{cases}
\end{equation}

The following example illustrates the above observation.


\emph{\textbf{Example 3}}. Given the arc-colored regular digraph of Fig.2, based on (2), we have
the following array \textbf{P}.
It can be seen that \textbf{P} is a $(12,12,8,10)$ PDA.
\begin{equation*}
\label{Eq:matrix1}
\textbf{P}=\bordermatrix{%
 & v_0 & v_1& v_2 & v_3& v_4 & v_5&v_6&v_7&v_8&v_9&v_{10}&v_{11}\cr
   v_0    &\ast & 1 & 9& \ast&6&\ast&\ast&\ast&4&\ast&\ast&\ast\cr
    v_1  &0 & \ast & \ast& 9&\ast&6&\ast&\ast&\ast&4&\ast&\ast\cr
     v_2   &8 & \ast & \ast& 1&\ast&\ast&6&\ast&\ast&\ast&4&\ast\cr
     v_3  &\ast & 8 & 0& \ast&\ast&\ast&\ast&6&\ast&\ast&\ast&4\cr
     v_4 &2 & \ast & \ast& \ast&\ast&1&9&\ast&7&\ast&\ast&\ast\cr
    v_5 &\ast & 2 & \ast& \ast&0&\ast&\ast&9&\ast&7&\ast&\ast\cr
    v_6&\ast & \ast & 2& \ast&8&\ast&\ast&1&\ast&\ast&7&\ast\cr
     v_7&\ast &\ast & \ast& 2&\ast&8&0&\ast&\ast&\ast&\ast&7\cr
      v_8&5 &\ast & \ast& \ast&3&\ast&\ast&\ast&\ast&1&9&\ast\cr
       v_9&\ast &5 & \ast&\ast&\ast&3&\ast&\ast&0&\ast&\ast&9\cr
        v_{10}&\ast &\ast &5& \ast&\ast&\ast&3&\ast&8&\ast&\ast&1\cr
         v_{11}&\ast &\ast & \ast& 5&\ast&\ast&\ast&3&\ast&8&0&\ast\cr
  }.
\end{equation*}

Based on the above investigation, the following theorem that describes the relationship between the injective arc coloring of a regular digraph and a PDA with the same number of rows and columns can be reached.

\emph{\textbf{Theorem 1}}. For any injective arc-colored regular digraph $D$ with the number of
vertices $K$ and indegree $K-Z$, if the arcs of $D$ can be colored by the colors $0,1,\ldots,S-1$,
the corresponding array \textbf{P} is a $(K,K,Z,S)$ PDA.

\begin{proof}
Given an injective arc coloring of regular digraph with the number of vertices $K$, one needs to show that the resulting array satisfies the definition of PDA. Since $D$ is a regular digraph, each vertex has the same indegree and outdegree. Let us assume that the indegree of each vertex is $K-Z$. Based on the entry rule of (2), each column of \textbf{P} has $Z$ $``\ast"$s. Furthermore, it is impossible for an entry to appear more than once in each row or each column. This is because any two arcs with the same head or tail receive the distinct colors due to its injective arc coloring. If there exist two distinct entries such that $\textbf{P}(v_{j_1},v_{k_1})=\textbf{P}(v_{j_2},v_{k_2})=s$, arcs
$(v_{j_1},v_{k_1})$ and $(v_{j_2},v_{k_2})$ are colored by $s$. Note that this coloring is an injective arc coloring and $D$ is a digraph without a directed self loop. This implies that $(v_{k_1},v_{j_2})$ and $(v_{k_2},v_{j_1})$ are not arcs of $D$. Since the regular digraph has reverse arcs, it can be seen that both $(v_{j_1},v_{k_2})$ and $(v_{j_2},v_{k_1})$ are also not arcs of $D$, i.e., $\textbf{P}(v_{j_1},v_{k_2})=\textbf{P}(v_{j_2},v_{k_1})=\ast$. Hence, the array \textbf{P} defined in (2) is a $(K,K,Z,S)$ PDA.
\end{proof}

This provides a new method to construct PDAs from the arc injective coloring of regular digraphs.
 It can be seen that a PDA characterized by {\it Theorem 1} can realize a coded caching scheme with a transmission rate of $R=\frac{S}{K}$. For fixed $K$ and $Z$, we would like to obtain a PDA scheme with the transmission rate that is as small as possible. This implies that the number of colors $S$ needed for an injective arc coloring of a regular digraph should be as small as possible.
 Therefore, it is important to properly construct a regular digraph and determine its optimal or near optimal number of injective arc coloring, which will be discussed in the next section.



\section{A New Construction of Regular Digraphs and Their Injective Arc Colorings}
This section defines a new class of regular digraphs and derives the upper bounds of their injective chromatic index. Given two vectors $\textbf{x}$ and $\textbf{y}$, the Hamming distance between \textbf{x} and \textbf{y} is defined as the number of coordinates that \textbf{x} and \textbf{y} differ, and denoted as $d_{\rm H}(\textbf{x},\textbf{y})$. 
Consequently, a regular digraph $\hat{D}$ with the number of vertices $p_0^{n_0}p_1^{n_1}\cdots p_{m-1}^{n_{m-1}}$ can be defined as follows.

\emph{\textbf{Definition 6}}. Given any $w,n_i,m\in \mathbb{N}^{+}$ and distinct positive integers $p_0, p_1, \ldots, p_{m-1}$ with $p_i\geq2$ and $w< n_0+n_1+\cdots+n_{m-1}$ for $i\in[0:m-1]$, the vertex set $\mathcal{V}$ and arc set $\mathcal{E}$ of $\hat{D}$ are defined as

 $\mathcal{V}=\{\textbf{x}=(\underbrace{x_0^{(n_0)}, x_1^{(n_0)}, \ldots, x_{n_0-1}^{(n_0)}}_{n_0}, \underbrace{x_{0}^{(n_1)}, x_{1}^{(n_1)}, \ldots, x_{n_1-1}^{(n_1)}}_{n_1}, \ldots, \underbrace{x_{0}^{(n_{m-1})}, x_{1}^{(n_{m-1})}, \ldots, x_{n_m-1}^{(n_{m-1})}}_{n_{m-1}})\mid \textbf{x}\in\mathbb{Z}_{p_0}^{n_0}\times\mathbb{Z}_{p_1}^{n_1}\times\cdots\times\mathbb{Z}_{p_{m-1}}^{n_{m-1}}\}$,\\
 and
  $$\mathcal{E}=\{(\textbf{x},\textbf{y}),(\textbf{y},\textbf{x})\mid d_{\rm H}(\textbf{x},\textbf{y})=w,\textbf{x},\textbf{y}\in\mathcal{V}\},$$
  respectively.




  In the following, we will present an upper bound on the injective chromatic index of $\hat{D}$. To do so, these arcs are partitioned into several disjoint subsets. A color is assigned to each of the arc partition such that the resulting coloring is an injective arc coloring.

Given any arc $(\textbf{x},\textbf{y})\in \mathcal{E}$, let $\mathcal{C}_{\textbf{x}-\textbf{y}}$ denote the set of the coordinates where two vectors \textbf{x} and \textbf{y} differ, i.e., $$\mathcal{C}_{\textbf{x}-\textbf{y}}=\{(j^{(i)}, n_i)\mid x_{j^{(i)}}^{(n_i)}\neq y_{j^{(i)}}^{(n_i)}, j^{(i)}\in[0:n_i-1], i\in[0: m-1]\}.$$ Let $\mathcal{C}_{\textbf{e}}$ denote the set of nonzero coordinates in \textbf{e}, where $\textbf{e}=\textbf{x}-\textbf{y}$ and $|\mathcal{C}_{\textbf{e}}|=w$.
Let $$\mathcal{T}_{\textbf{e}}=\{\textbf{t}\mid \textbf{t}\in\mathbb{Z}_{p_{i_0}}\times\mathbb{Z}_{p_{i_1}}\times\cdots\times\mathbb{Z}_{p_{i_{w-1}}}, i_0 \leq i_1\leq\cdots\leq i_{w-1}\}$$ denote a set of vectors with length $w$, where non-negative integers  $i_0, i_1, \ldots, i_{w-1}$ are determined by $\mathcal{C}_\textbf{e}=\{(j_{\alpha_0}^{(i_0)}, n_{i_0}), (j_{\alpha_1}^{(i_1)}, n_{i_1}),\ldots, (j_{\alpha_{w-1}}^{(i_{w-1})}, n_{i_{w-1}})\}$ and $j_{\alpha_{u}}^{(i_u)}\in[0:n_{i_{u}}-1]$ for $u\in[0:w-1]$.
The arc set $\mathcal{E}$ of $\hat{D}$ can be partitioned as
\begin{equation}
\mathcal{E}=\bigcup\limits_{\substack{\textbf{e}\in\{\textbf{x}-\textbf{y}\mid (\textbf{x},\textbf{y})\in\mathcal{E}\}}}\mathcal{E}_{\textbf{e}}=\bigcup\limits_{\substack{\textbf{e}\in\{\textbf{x}-\textbf{y}\mid (\textbf{x},\textbf{y})\in\mathcal{E}\}}}\bigcup\limits_{\textbf{t}\in\mathcal{T}_{\textbf{e}}}\mathcal{E}_{\textbf{e},\textbf{t}},
\end{equation}
where \[\mathcal{E}_{\textbf{e}}=\{(\textbf{x},\textbf{y})\mid (\textbf{x},\textbf{y})\in\mathcal{E}, \textbf{x}-\textbf{y}=\textbf{e}\},\] \[\mathcal{E}_{\textbf{e},\textbf{t}}=\{(\textbf{x},\textbf{y})\mid (\textbf{x},\textbf{y})\in\mathcal{E}, \textbf{x}-\textbf{y}=\textbf{e}, \textbf{y}|_{\mathcal{C}_{\textbf{e}}}=\textbf{t}\}\] and the computations in the $n_i$th block are performed under modulo $p_i$ with $i\in[0:m-1]$.

The above arc partition leads to the following result.

\emph{\textbf{Proposition 1}}. The assignment of a distinct color for each subset $\mathcal{E}_{\textbf{e},\textbf{t}}$ forms an injective arc coloring of $\hat{D}$.


\begin{proof}
Let us consider $\mathcal{E}_{\textbf{e},\textbf{t}}$ for any \textbf{e} and \textbf{t} defined above. It can be seen that $|\mathcal{E}_{\textbf{e}, \textbf{t}}|>1$. Based on (3), for any two distinct arcs $(\textbf{x}_1, \textbf{y}_1), (\textbf{x}_2, \textbf{y}_2)\in\mathcal{E}_{\textbf{e}, \textbf{t}}$, we have

\begin{equation}
        \begin{cases}
         \textbf{x}_1-\textbf{y}_1=\textbf{x}_2-\textbf{y}_2=\textbf{e},\\
         \textbf{y}_{1}|_{\mathcal{C}_\textbf{e}}=
         \textbf{y}_{2}|_{\mathcal{C}_\textbf{e}}=\textbf{t}.
 \end{cases}
\end{equation}
This implies that $\textbf{x}_{1}|_{\mathcal{C}_\textbf{e}}=\textbf{x}_{2}|_{\mathcal{C}_\textbf{e}}$. Hence, $d_{\rm H}(\textbf{x}_1, \textbf{y}_2)\geq d_{\rm H}(\textbf{x}_1|_{\mathcal{C}_\textbf{e}}, \textbf{y}_2|_{\mathcal{C}_\textbf{e}})=d_{\rm H}(\textbf{x}_2|_{\mathcal{C}_\textbf{e}}, \textbf{y}_2|_{\mathcal{C}_\textbf{e}})=w$. Let $\mathcal{C}_\textbf{e}'$ denote the set of zero coordinates in \textbf{e}. If $d_{\rm H}(\textbf{x}_1, \textbf{y}_2)=w$, we have $\textbf{x}_{1}|_{\mathcal{C}_\textbf{e}'}=\textbf{y}_{2}|_{\mathcal{C}_\textbf{e}'}$ since
$d_{\rm H}(\textbf{x}_1, \textbf{y}_2)=d_{\rm H}(\textbf{x}_1|_{\mathcal{C}_\textbf{e}}, \textbf{y}_2|_{\mathcal{C}_\textbf{e}})+d_{\rm H}(\textbf{x}_1|_{\mathcal{C}_{\textbf{e}'}}, \textbf{y}_2|_{\mathcal{C}_{\textbf{e}'}})=d_{\rm H}(\textbf{x}_1|_{\mathcal{C}_\textbf{e}}, \textbf{y}_1|_{\mathcal{C}_\textbf{e}})+d_{\rm H}(\textbf{x}_1|_{\mathcal{C}_{\textbf{e}'}}, \textbf{y}_2|_{\mathcal{C}_{\textbf{e}'}})=w+d_{\rm H}(\textbf{x}_1|_{\mathcal{C}_{\textbf{e}'}}, \textbf{y}_2|_{\mathcal{C}_{\textbf{e}'}})=w$.
 Hence, $\textbf{x}_{1}|_{\mathcal{C}_\textbf{e}'}=\textbf{x}_{2}|_{\mathcal{C}_\textbf{e}'}=
\textbf{y}_{1}|_{\mathcal{C}_\textbf{e}'}=\textbf{y}_{2}|_{\mathcal{C}_\textbf{e}'}$ and we have $\textbf{x}_1=\textbf{x}_2$ and $\textbf{y}_1=\textbf{y}_2$, which contradicts the hypothesis. Therefore, $d_{\rm H}(\textbf{x}_1, \textbf{y}_2)>w$. Similarly, we also have $d_{\rm H}(\textbf{x}_2, \textbf{y}_1)>w$. This implies that $(\textbf{x}_1, \textbf{y}_2)$, $(\textbf{y}_2, \textbf{x}_1)$, $(\textbf{x}_2, \textbf{y}_1)$ and $(\textbf{y}_1, \textbf{x}_2)$ are not arcs of $\hat{D}$. Therefore, the assignment of a distinct color for each subset $\mathcal{E}_{\textbf{e},\textbf{t}}$ forms an injective arc coloring.
\end{proof}
\begin{figure}[H]
\begin{center}
\includegraphics[scale=0.54]{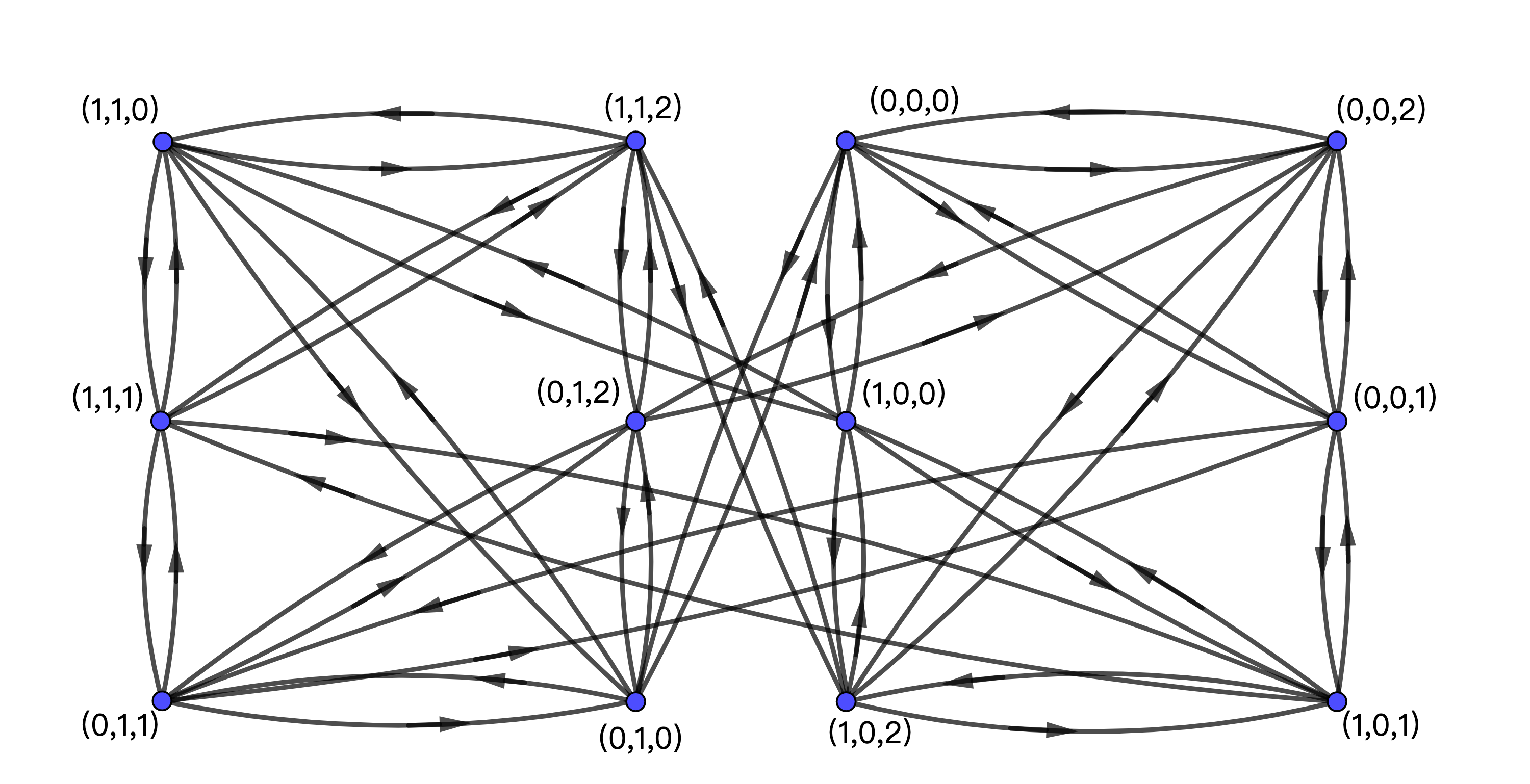}
\caption{\label{2}4-regular digraph.}
 \end{center}
 \end{figure}
The following example illustrates the above property.

\emph{\textbf{Example 4}}. Given $m=2, w=1, p_0=2, p_1=3$, $n_0=2$ and $n_1=1$, based on {\it Definition 6}, a regular digraph shown in Fig.3 can be obtained.
Based on (3), its arcs can be partitioned as

$\mathcal{E}_{(0,1,0),(0)}=\{((0,1,0),(0,0,0)),((0,1,1),(0,0,1)), ((0,1,2),(0,0,2)), ((1,1,0),(1,0,0)), $\\$((1,1,1),(1,0,1)),((1,1,2),(1,0,2))\};$

$\mathcal{E}_{(0,1,0),(1)}=\{((0,0,0),(0,1,0)),((0,0,1),(0,1,1)), ((0,0,2),(0,1,2)), ((1,0,0),(1,1,0)), $\\$((1,0,1),(1,1,1)),((1,0,2),(1,1,2))\};$

$\mathcal{E}_{(0,0,1),(0)}=\{((0,0,1),(0,0,0)),((1,0,1),(1,0,0)), ((0,1,1),(0,1,0)), ((1,1,1),(1,1,0))\};$

$\mathcal{E}_{(0,0,1),(1)}=\{((0,0,2),(0,0,1)),((1,0,2),(1,0,1)), ((0,1,2),(0,1,1)), ((1,1,2),(1,1,1))\};$

$\mathcal{E}_{(0,0,1),(2)}=\{((0,0,0),(0,0,2)),((1,0,0),(1,0,2)), ((0,1,0),(0,1,2)), ((1,1,0),(1,1,2))\};$

$\mathcal{E}_{(0,0,2),(0)}=\{((0,0,2),(0,0,0)),((1,0,2),(1,0,0)), ((0,1,2),(0,1,0)), ((1,1,2),(1,1,0))\};$

$\mathcal{E}_{(0,0,2),(1)}=\{((0,0,0),(0,0,1)),((1,0,0),(1,0,1)), ((0,1,0),(0,1,1)), ((1,1,0),(1,1,1))\};$

$\mathcal{E}_{(0,0,2),(2)}=\{((0,0,1),(0,0,2)),((1,0,1),(1,0,2)), ((0,1,1),(0,1,2)), ((1,1,1),(1,1,2))\};$

$\mathcal{E}_{(1,0,0),(0)}=\{((1,0,0),(0,0,0)),((1,1,0),(0,1,0)), ((1,0,1),(0,0,1)), ((1,1,1),(0,1,1)), $\\$((1,0,2),(0,0,2)), ((1,1,2),(0,1,2))\};$

$\mathcal{E}_{(1,0,0),(1)}=\{((0,0,0),(1,0,0)),((0,1,0),(1,1,0)), ((0,0,1),(1,0,1)), ((0,1,1),(1,1,1)), $\\$((0,0,2),(1,0,2)), ((0,1,2),(1,1,2))\};$


If each partitioned subset is assigned with a distinct color,
a regular digraph with the number of injective arc coloring of 10, which is shown in Fig.2, can be obtained.

Based on {\it Proposition 1}, the injective chromatic index of $\hat{D}$ can be determined by the following theorem.

\emph{\textbf{Theorem 2}}. For the digraph $\hat{D}$, $\chi_{i}(\hat{D})\leq\sum_{\textbf{e}\in\{\textbf{x}-\textbf{y}\mid (\textbf{x},\textbf{y})\in\mathcal{E}\}}p_{i_0}p_{i_1}\cdots p_{i_{w-1}},$ where integers $i_0,i_1, \ldots, i_{w-1}$ are determined by $\mathcal{C}_\textbf{e}=\{(j_{\alpha_0}^{(i_0)}, n_{i_0}), (j_{\alpha_1}^{(i_1)}, n_{i_1}),\ldots, (j_{\alpha_{w-1}}^{(i_{w-1})}, n_{i_{w-1}})\}$.

\begin{proof}
Given any $\textbf{e}\in\{\textbf{x}-\textbf{y}\mid(\textbf{x},\textbf{y})\in\mathcal{E}\}$, based on (3), it can be seen that the arc set $\mathcal{E}_{\textbf{e}}=\{(\textbf{x},\textbf{y})\mid \textbf{x}-\textbf{y}=\textbf{e}, (\textbf{x},\textbf{y})\in\mathcal{E}\}$ is partitioned into $p_{i_0}p_{i_1}\cdots p_{i_{w-1}}$ subsets since $|\mathcal{T}_{\textbf{e}}|=p_{i_0}p_{i_1}\cdots p_{i_{w-1}}$. Hence, the total number of partitioned subsets is
$$\sum_{\textbf{e}\in\{\textbf{x}-\textbf{y}\mid(\textbf{x},\textbf{y})\in\mathcal{E}\}}|\mathcal{T}_{\textbf{e}}|=      \sum_{\textbf{e}\in\{\textbf{x}-\textbf{y}\mid(\textbf{x},\textbf{y})\in\mathcal{E}\}}p_{i_0}p_{i_1}\cdots p_{i_{w-1}},$$ i.e., the number of colors that allow an injective arc coloring of $\hat{D}$ is less than or equal to $$\sum_{\textbf{e}\in\{\textbf{x}-\textbf{y}\mid(\textbf{x},\textbf{y})\in\mathcal{E}\}}p_{i_0}p_{i_1}\cdots p_{i_{w-1}}.$$
\end{proof}

 In particular, let $\overline{D}$ denote a regular digraph defined in {\it Definition 6} with parameter $m=1$.
An upper bound on the injective chromatic index of $\overline{D}$ can be obtained as follows, which can be viewed as a special case of {\it Theorem 2}.

\emph{\textbf{Corollary 1}}. For the digraph $\overline{D}$, $\chi_{i}(\overline{D})\leq{n_0\choose w}p_{0}^{w}(p_{0}-1)^{w}.$
\begin{proof}
Let $\overline{D}$ denote a digraph with vertex set $\mathcal{V}=\{\textbf{x}=(x_0, x_1, \ldots, x_{n_{0}-1})\mid \textbf{x}\in\mathbb{Z}_{p_{0}}^{n_0}\}$ and arc set $\mathcal{E}=\{(\textbf{x},\textbf{y}),(\textbf{y},\textbf{x})\mid d_{\rm H}(\textbf{x},\textbf{y})=w, \textbf{x},\textbf{y}\in\mathcal{V}\}$. Note that the cardinality of set $\{\textbf{x}-\textbf{y}\mid(\textbf{x},\textbf{y})\in\mathcal{E}\}$ is ${n_0\choose w}(p_0-1)^{w}$.
 Based on (3), it can be seen that each $\mathcal{E}_{\textbf{e}}=\{(\textbf{x},\textbf{y})\mid \textbf{x}-\textbf{y}=\textbf{e}, (\textbf{x},\textbf{y})\in\mathcal{E}\}$ is partitioned into $p_{0}^{w}$ subsets since $|\mathcal{T}_{\textbf{e}}|=p_0^{w}$. This implies that the total number of partitioned subsets is ${n_0\choose w}p_{0}^{w}(p_{0}-1)^{w}$, i.e., the number of colors that allow an injective arc coloring of $\overline{D}$ is less than or equal to ${n_0\choose w}p_{0}^{w}(p_0-1)^{w}$.
\end{proof}

In fact, given the graph $\overline{D}$ with parameters $n_0, w$ and $p_0$ such that $p_0=2$ and $n_0\leq 2w-1$, the upper bound on injective chromatic index described in {\it Corollary 1} can be further improved by vertex coloring. A vertex coloring of a graph $G$ is an assignment of colors to the vertices of $G$ with one color for each vertex, so that the adjacent vertices are colored differently. The smallest number of
colors needed in a proper vertex coloring of graph $G$ is called the chromatic index of $G$, denoted by $\chi(G)$.
 It is well known that $\chi(G)\leq 1+\triangle(G)$ \cite{564}, where $\triangle(G)$ is the maximal degree of $G$. In order to improve the upper bound on injective chromatic index of $\overline{D}$, we need the following lemma.

\emph{\textbf{Lemma 2}}. Given the digraph $\overline{D}$ with parameters $n_0, w$ and $p_0$ such that $p_0=2$ and $n_0\leq 2w-1$, let the arcs of $\overline{D}$ be partitioned as
\begin{equation}
\mathcal{E}=\bigcup_{\textbf{e}\in\{\textbf{x}-\textbf{y}\mid(\textbf{x},\textbf{y})\in\mathcal{E}
\}}\mathcal{E}_{\textbf{e}}=\bigcup_{\textbf{e}\in\{\textbf{x}-\textbf{y}\mid(\textbf{x},\textbf{y})\in\mathcal{E}
\}}\bigcup
_{i=0}^{g_{\textbf{e}-1}}\mathcal{G}_{\textbf{e},\mathcal{D}_i},
 \end{equation}
 where $\mathcal{G}_{\textbf{e},\mathcal{D}_i}=\{\mathcal{E}_{\textbf{e},\textbf{t}_i}\mid \textbf{t}_i\in\mathcal{D}_i\}$, $\mathcal{D}_i$ is a subset of $\mathbb{Z}_2^{w}$ such that $d_{\rm H}(\textbf{t}_i,\textbf{t}_j)\geq n_0-w+1$ for any $\textbf{t}_i,\textbf{t}_j\in\mathcal{D}_i$, and $\mathbb{Z}_2^{w}=\mathcal{D}_0\cup \mathcal{D}_1\cup \cdots \cup \mathcal{D}_{g_{\textbf{e}}-1}$. Consequently, assigning a distinct color for each subset $\mathcal{G}_{\textbf{e},\mathcal{D}_i}$ forms an injective arc coloring.

\begin{proof} It is sufficient to prove that any two arcs at a distance of exactly one in a directed path or in a common directed triangle receive distinct colors. Without loss of generality, for any two arcs $(\textbf{x}_1,\textbf{y}_1),(\textbf{x}_2,\textbf{y}_2)\in\mathcal{G}_{\textbf{e},\mathcal{D}_i}$, if $(\textbf{x}_1,\textbf{y}_1),(\textbf{x}_2,\textbf{y}_2)\in\mathcal{E}_{\textbf{e},\textbf{t}_i}$, based on {\it Proposition 1}, they can be assigned with the same color.
Now let us consider the case $(\textbf{x}_1,\textbf{y}_1)\in\mathcal{E}_{\textbf{e},\textbf{t}_i}, (\textbf{x}_2,\textbf{y}_2)\in\mathcal{E}_{\textbf{e},\textbf{t}_j}$ and $\textbf{t}_i\neq\textbf{t}_j$. One needs to prove that arcs $(\textbf{x}_1,\textbf{y}_1)$ and $(\textbf{x}_2,\textbf{y}_2)$ can also be assigned with the same color. Based on (5),
we obtain
$\textbf{t}_i=\textbf{y}_1|_{\mathcal{C}_{\textbf{e}}}$ and $\textbf{t}_j=\textbf{y}_2|_{\mathcal{C}_{\textbf{e}}}$. Note that $d_{\rm H}(\textbf{t}_i,\textbf{t}_j)\geq n_0-w+1$. There must exist a set $\mathcal{A}\subseteq \mathcal{C}_{\textbf{e}}$ with $|\mathcal{A}|\geq n_0-w+1$ such that $\textbf{y}_1|_s\neq\textbf{y}_2|_s$ for any $s\in\mathcal{A}$. Furthermore, it can be seen that $\textbf{x}_1|_s=\textbf{y}_2|_s$ and $\textbf{x}_2|_s=\textbf{y}_1|_s$ for $s\in\mathcal{A}$, since $\textbf{x}_1-\textbf{y}_1=\textbf{x}_2-\textbf{y}_2=\textbf{e}$ and $p_0=2$. This implies that $d_{\rm H}(\textbf{x}_1,\textbf{y}_2)=d_{\rm H}(\textbf{x}_1|_{\mathcal{C}_{\textbf{e}}}, \textbf{y}_2|_{\mathcal{C}_{\textbf{e}}})+d_{\rm H}(\textbf{x}_1|_{[0:n_0-1]\backslash \mathcal{C}_{\textbf{e}}}, \textbf{y}_2|_{[0:n_0-1]\backslash \mathcal{C}_{\textbf{e}}})=d_{\rm H}(\textbf{x}_1|_{\mathcal{C}_{\textbf{e}}\backslash\mathcal{A}}, \textbf{y}_2|_{\mathcal{C}_{\textbf{e}}\backslash\mathcal{A}})+d_{\rm H}(\textbf{x}_1|_{[0:n_0-1]\backslash \mathcal{C}_{\textbf{e}}}, \textbf{y}_2|_{[0:n_0-1]\backslash \mathcal{C}_{\textbf{e}}})\leq w-(n_0-w+1)+n_0-w=w-1$. Therefore, $(\textbf{x}_1, \textbf{y}_2),(\textbf{y}_2, \textbf{x}_1)\notin\mathcal{E}$. Following a similar proof manner, it can also be concluded that
$(\textbf{x}_2, \textbf{y}_1), (\textbf{y}_1, \textbf{x}_2)\notin\mathcal{E}$. Therefore, assigning a distinct color for each subset
$\mathcal{G}_{\textbf{e},\mathcal{D}_i}$ forms an injective arc coloring.
\end{proof}


\emph{\textbf{Theorem 3}}. Given the digraph $\overline{D}$ with parameters $n_0, w$ and $p_0$ such that $p_0=2$ and $n_0\leq 2w-1$, we have

\begin{equation*}
       \chi_{i}(\overline{D})\leq \begin{cases}
         {n_0\choose w}2^{w-1}, \text{if $n_0=2w-1$};\\
         {n_0\choose w}(1+\sum_{i=1}^{n_0-w}{w\choose i}), \text{if $n_0<2w-1$}.
 \end{cases}
\end{equation*}


\begin{proof}
Let $\overline{D}$ denote a digraph with vertex set $\mathcal{V}=\{\textbf{x}=(x_0, x_1, \ldots, x_{n_0-1})\mid \textbf{x}\in\mathbb{Z}_{2}^{n_0}\}$ and arc set $\mathcal{E}=\{(\textbf{x},\textbf{y}),(\textbf{y},\textbf{x})\mid d_{\rm H}(\textbf{x},\textbf{y})=w, \textbf{x},\textbf{y}\in\mathcal{V}\}$. Note that the arc partition of $\overline{D}$ is
$\mathcal{E}=\bigcup_{\substack{\textbf{e}\in\{\textbf{x}-\textbf{y}\mid (\textbf{x},\textbf{y})\in\mathcal{E}\}}}\bigcup_{\textbf{t}\in\mathbb{Z}_2^{w}}\mathcal{E}_{\textbf{e},\textbf{t}}$,
where $\mathcal{E}_{\textbf{e},\textbf{t}}=\{(\textbf{x},\textbf{y})\mid (\textbf{x},\textbf{y})\in\mathcal{E}, \textbf{x}-\textbf{y}=\textbf{e}, \textbf{y}|_{\mathcal{C}_{\textbf{e}}}=\textbf{t}\}$. If $n_0<2w-1$, we can merge two subsets $\mathcal{E}_{\textbf{e},\textbf{t}_i}$ and $\mathcal{E}_{\textbf{e},\textbf{t}_j}$ for $\textbf{t}_i, \textbf{t}_j\in\mathbb{Z}_{2}^{w}$ under the condition of $d_{\rm H}(\textbf{t}_i,\textbf{t}_j)\geq n_0-w+1$.

Now let us determine the number of colors that allow an injective arc coloring to $\overline{D}$. If $n_0=2w-1$, we have $w=n_0-w+1\leq d_{\rm H}(\textbf{t}_i,\textbf{t}_j)\leq w$.
This implies that for any $\textbf{t}_i, \textbf{t}_j\in\mathbb{Z}_{2}^{w}$, two subsets $\mathcal{E}_{\textbf{e},\textbf{t}_i}$ and $\mathcal{E}_{\textbf{e},\textbf{t}_j}$ can be merged if and only if $d_{\rm H}(\textbf{t}_i,\textbf{t}_j)=w$. That says for any $\textbf{t}_i, \textbf{t}_j\in\mathbb{Z}_{2}^{w}$, two subsets $\mathcal{E}_{\textbf{e},\textbf{t}_i}$ and $\mathcal{E}_{\textbf{e},\textbf{t}_j}$ can be merged if and only if $\textbf{t}_i+\textbf{t}_j=\textbf{1}$. Therefore, each $\mathcal{E}_\textbf{e}$ is partitioned into $2^{w-1}$ subsets, where $\mathcal{E}_{\textbf{e}}=\{(\textbf{x},\textbf{y})\mid \textbf{x}-\textbf{y}=\textbf{e}, (\textbf{x},\textbf{y})\in \mathcal{E}\}$. Note that the cardinality of set $\{\textbf{x}-\textbf{y}\mid(\textbf{x},\textbf{y})\in \mathcal{E}\}$ is ${n_0\choose w}$. Hence, the total number of partitioned subsets is ${n_0\choose w}2^{w-1}$. Based on {\it Lemma 2}, it can be seen that the number of colors that allow an injective arc coloring of $\overline{D}$ is less than or equal to ${n_0\choose w}2^{w-1}$.

If $n_0<2w-1$, we can determine the injective arc coloring by using the vertex coloring of graph. Define a graph $G$ with
vertex set $\mathcal{V}=[0: 1]^{w}$ such that there exists an edge between two vertices $\textbf{t}_i$ and $\textbf{t}_j$ in $\mathcal{V}$ if
and only if $d_{\rm H}(\textbf{t}_i,\textbf{t}_j)\leq n_0-w$. Given any vertex $\textbf{t}_i\in\mathcal{V}$, the number of vertices of $G$ that are adjacent to $\textbf{t}_i$ is $\sum_{i=1}^{n_0-w}{w\choose i}$, i.e., $G$ is a $\sum_{i=1}^{n_0-w}{w\choose i}$-regular graph.
As a result, one color can be assigned to each vertex of $G$ such that all adjacent vertices receive distinct colors. This indicates that $d_{\rm H}(\textbf{t}_i,\textbf{t}_j)\geq n_0-w+1$ for any two vertices $\textbf{t}_i$ and $\textbf{t}_j$ with the same color, i.e., two subsets $\mathcal{E}_{\textbf{e},\textbf{t}_i}$ and $\mathcal{E}_{\textbf{e},\textbf{t}_j}$ can be merged. Based on the upper bound of chromatic number of $G$, it can be seen that each $\mathcal{E}_\textbf{e}$ is partitioned into $\chi(G)$ subsets, where $\chi(G)\leq 1+\sum_{i=1}^{n_0-w}{w\choose i}$. This implies that the total number of partitioned subsets is less than or equal to ${n_{0}\choose w}(1+\sum_{i=1}^{n_0-w}{w\choose i})$. Based on {\it Lemma 2}, it can be seen that the number of colors that allow an injective arc coloring of $\overline{D}$
is less than or equal to ${n_0\choose w}(1+\sum_{i=1}^{n_0-w}{w\choose i}).$
\end{proof}
\section{The New PDA Schemes}
Based on the upper bounds on the injective chromatic indices of regular digraphs, some new coded caching schemes that support a flexible number of users and with a linear subpacketization level can be further proposed.

\subsection{New PDA Constructions via Injective Arc Coloring of Regular Digraphs}
The injective arc coloring of regular digraphs is employed for PDA design. In fact, if each colored edge of a regular graph is replaced by two reverse arcs with the same color, the strong edge coloring of regular graphs can be viewed as a special injective arc coloring of regular digraphs, as illustrated by Fig.4.
\begin{figure}[H]
 \centering
\subfigure[]{
\begin{minipage}{7.6cm}
\centering
\includegraphics[scale=0.71]{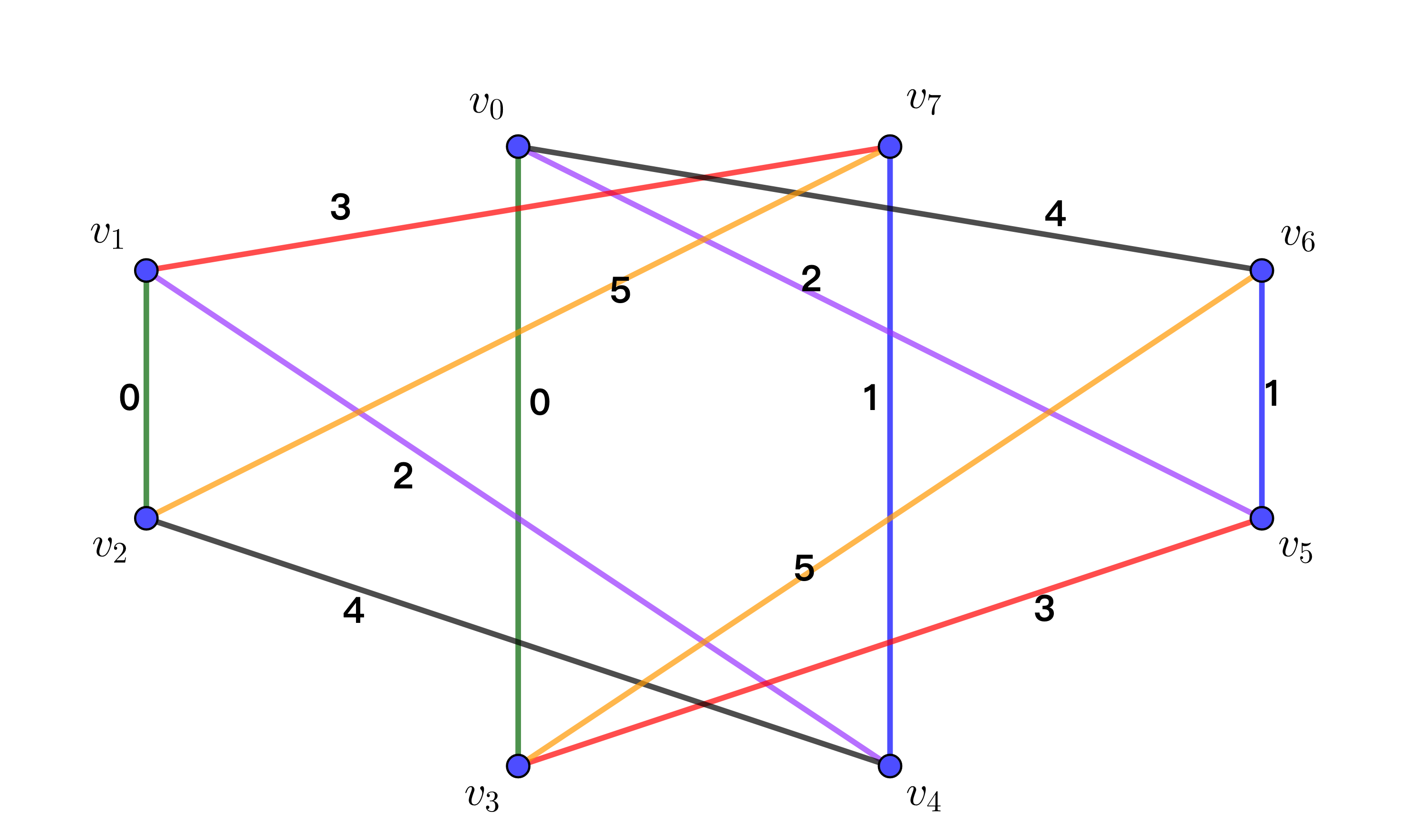}
\end{minipage}%
}%
\subfigure[]{
\begin{minipage}{7.6cm}
\centering
\includegraphics[scale=0.7]{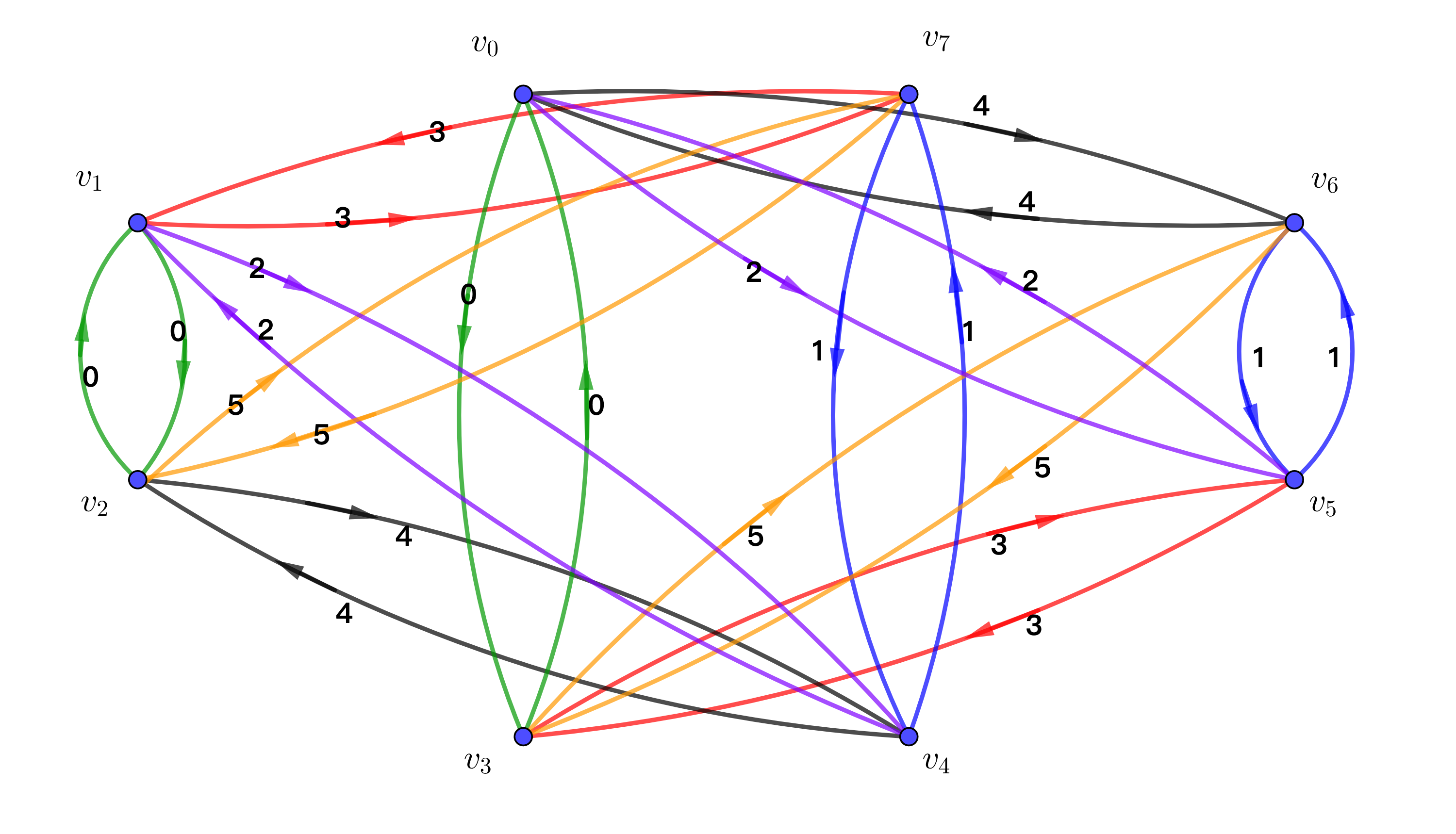}
\end{minipage}
}
\caption{(a) 3-regular graph with the number of strong edge coloring of 6; (b) 3-regular digraph with the number of injective arc coloring of 6.}
\end{figure}




Based on {\it Theorem 1}, some new coded caching schemes with a linear subpacketization level can be obtained from some existing strong edge coloring of regular graphs. A unitary Cayley graph is a graph with vertex set $\mathbb{Z}_n$ and edge set $\mathcal{E}=\{\{i,j\}\mid$ gcd$(i-j,n)=1,i,j\in\mathbb{Z}_n\}$, where gcd$(i-j,n)$ denotes the greatest common divisor between integers $i-j$ and $n$. It can be observed that a unitary Cayley graph is a regular graph with degree $\psi(n)$, where $\psi(n)$ denotes the Euler function, i.e., the number of integers that are less than $n$ and relatively prime to $n$. The strong chromatic index of Cayley graphs is characterized as follows.


 \emph{\textbf{Lemma 3}} \cite{556}. Given any positive integer $n=p_0^{n_0}p_1^{n_1}\cdots p_{m-1}^{n_{m-1}}$ with prime factor $p_i\geq2$ for $i\in[0: m-1]$, the strong chromatic index of unitary Cayley graphs is $\frac{n\psi(n)}{2^{m}}$.

Based on {\it Theorem 1} and {\it Lemma 3}, we have the following result.


\emph{\textbf{Theorem 4}}. Given any positive integer $n=p_0^{n_0}p_1^{n_1}\cdots p_{m-1}^{n_{m-1}}$ with prime factor $p_i\geq2$ for $i\in [1:m-1]$, there always exists an $(n, n, n-\psi(n), \frac{n\psi(n)}{2^{m}})$ PDA
which yields an $n$-division $(n, M, N)$ coded caching scheme with a memory ratio of $$\frac{M}{N}=1-\frac{\psi(n)}{n},$$ and a transmission rate of $$R=\frac{\psi(n)}{2^{m}}.$$

Integrating {\it Theorems 1} and {\it 2}, a new coded caching scheme that is characterized by the following theorem can be obtained.


\emph{\textbf{Theorem 5}}. Given any $w,n_i,m\in \mathbb{N}^{+}$ and distinct positive integers $p_0, p_1, \ldots, p_{m-1}$ with $p_i\geq2$ and $w< n_0+n_1+\cdots+n_{m-1}$ for $i\in[0:m-1]$, there exists a $(p_{0}^{n_0}p_{1}^{n_1}\cdots p_{m-1}^{n_{m-1}}$, $p_{0}^{n_0}p_{1}^{n_1}\cdots p_{m-1}^{n_{m-1}}$, $p_{0}^{n_0}p_{1}^{n_1}\cdots p_{m-1}^{n_{m-1}}-\sum_{\substack{\mathcal{A}\subseteq\mathcal{X},|\mathcal{A}|=w}}\prod_{p^{(\beta)}_{\alpha}\in\mathcal{A}}(p_{\alpha}-1)$, $\sum_{\textbf{e}\in\{\textbf{x}-\textbf{y}\mid(\textbf{x},\textbf{y})\in \mathcal{E}\}}p_{i_0}p_{i_1}\cdots p_{i_{w-1}}$) PDA which yields a $p_{0}^{n_0}p_{1}^{n_1}\cdots p_{m-1}^{n_{m-1}}$-division $(p_{0}^{n_0}p_{1}^{n_1}\cdots p_{m-1}^{n_{m-1}}, M, N)$ coded
caching scheme with a memory ratio of
$$\frac{M}{N}=1-\frac{\sum_{\substack{\mathcal{A}\subseteq\mathcal{X},|\mathcal{A}|=w}}\prod_{p^{(\beta)}_{\alpha}\in\mathcal{A}}
(p_{\alpha}-1)}{p_{0}^{n_0}p_{1}^{n_1}\cdots p_{m-1}^{n_{m-1}}},$$ and a transmission rate of
$$R=\frac{\sum_{\textbf{e}\in\{\textbf{x}-\textbf{y}\mid (\textbf{x},\textbf{y})\in \mathcal{E}\}}p_{i_0}p_{i_1}\cdots p_{i_{w-1}}} {p_{0}^{n_0}p_{1}^{n_1}\cdots p_{m-1}^{n_{m-1}}},$$ where
$$\mathcal{X}=\{\underbrace{p^{(0)}_0, p^{(1)}_0, \ldots, p^{(n_0-1)}_0}_{n_0}, \underbrace{p^{(0)}_1, p^{(1)}_1, \ldots, p^{(n_1-1)}_1}_{n_1}, \ldots, \underbrace{p^{(0)}_{m-1}, p^{(1)}_{m-1}, \ldots, p^{(n_{m-1}-1)}_{m-1}}_{n_{m-1}}\},$$
and integers $i_0, i_1, \ldots, i_{w-1}$ are determined by $\mathcal{C}_\textbf{e}=\{(j_{\alpha_0}^{(i_0)}, n_{i_0}), (j_{\alpha_1}^{(i_1)} n_{i_1}),\ldots, (j_{\alpha_{w-1}}^{(i_{w-1})}, n_{i_{w-1}})\}$.
\begin{proof}
Let $\hat{D}$ denote a regular digraph defined in {\it Definition 6}. It can be seen that the number of vertices of $\hat{D}$ is $|\mathcal{V}|=p_{0}^{n_0}p_{1}^{n_1}\ldots p_{m-1}^{n_{m-1}}$.
Given any vertex $\textbf{y}\in \mathcal{V}$, it can be seen that the number of vertices $\textbf{x}\in \mathcal{V}$ such that $d_{\rm H}(\textbf{x}, \textbf{y})=w$ is
$$\sum_{\substack{\mathcal{A}\subseteq\mathcal{X},|\mathcal{A}|=w}}\prod_{p^{(\beta)}_{\alpha}\in\mathcal{A}}(p_{\alpha}-1).$$
Therefore, $\hat{D}$ is a regular digraph with both the indegree and outdegree being

$$\sum_{\substack{\mathcal{A}\subseteq\mathcal{X},|\mathcal{A}|=w}}\prod_{p^{(\beta)}_{\alpha}\in\mathcal{A}}(p_{\alpha}-1).$$ Together with the results of {\it Theorems 1} and {\it 2}, the conclusion can be reached.
\end{proof}

%

Based on {\it Theorem 1} and {\it Corollary 1}, the following corollary can be obtained, which can be seen as a special case of {\it Theorem 5}.


\emph{\textbf{Corollary 2}}. Given any $n_0, w, p_0\in\mathbb{N}^{+}$ with $p_0\geq2$ and $w<n$, there always exists a $(p_{0}^{n_0}, p_{0}^{n_0}, p_{0}^{n_0}-{n_0\choose w}(p_0-1)^{w}, {n_0\choose w}p_{0}^{w}(p_0-1)^{w})$ PDA
which yields a $p_{0}^{n_0}$-division $(p_{0}^{n_0}, M, N)$ coded caching scheme with a memory ratio of
$$\frac{M}{N}=1-\frac{{n_0\choose w}(p_0-1)^{w}}{p_{0}^{n_0}},$$ and a transmission rate of $$R=\frac{{n_0\choose w}(p_0-1)^{w}}{p_{0}^{n_0-w}}.$$
\begin{proof}
Let $\overline{D}$ denote a regular digraph defined above with vertex set $\mathcal{V}$ and arc set $\mathcal{E}$. Given any vertex $\textbf{y}\in \mathcal{V}$, the number of vertices $\textbf{x}\in \mathcal{V}$ such that $d_{\rm H}(\textbf{x}, \textbf{y})=w$ is ${n_0\choose w}(p_0-1)^{w}$. This implies that both the outdegree and indegree of each vertex are
${n_0\choose w}(p_0-1)^{w}$. Based on the results of {\it Theorem 1} and {\it Corollary 1}, the conclusion can be reached.
\end{proof}

The following result is a straightforward application of {\it Theorems 1} and {\it 2}.

\emph{\textbf{Corollary 3}}. Given any $n_0, w, p_0\in\mathbb{N}^{+}$ with $p_0=2$ and $w<n_0$, there always exists a $2^{n_0}$-division $(2^{n_0}, M, N)$ coded caching scheme with a memory ratio of $$\frac{M}{N}=1-\frac{{n_0\choose w}}{2^{n_0}},$$
and a transmission rate of
\begin{equation}
       R= \begin{cases}
         \frac{{n_0\choose w}}{2^{n_0-w+1}}, \text{if $n_0=2w-1$};\\
         \frac{{n_0\choose w}(1+\sum_{i=1}^{n_0-w}{w\choose i})}{2^{n_0}}, \text{if $n_0<2w-1$}.
 \end{cases}
\end{equation}

 It can be seen that the coded caching schemes characterized in {\it Theorem 5}, {\it Corollaries 2} and {\it 3} require a high memory ratio, even though they all exhibit a linear subpacketization level. In the following subsection, we will show that the memory ratios and subpacketization levels of the schemes in {\it Theorem 5} and {\it Corollary 2} can be further reduced by using the MDS code in the placement phase.
\subsection{New PDA Schemes with Coded Placement}
In a PDA, a $``\ast"$ is called useless, if it is not contained in any subarray shown in C3-(b) of {\it Definition 1}. This indicates these useless $``\ast"$s cannot generate multicasting opportunities in the delivery phase, i.e., they have no contributions in reducing the transmission rate of a coded caching scheme realized by the PDA and they result in both a high memory ratio and a high subpacketization level. Therefore, if each column of a $(K, F, Z, S)$ PDA has $Z'$ useless $``\ast"$s, we can obtain a new coded caching scheme with a smaller memory ratio and subpacketization level by deleting these useless $``\ast"$s and using an $[F,F-Z']_q$ MDS code that is defined in a finite field of size $q$. For the detailed implementation method, interested readers can refer to \cite{24}.


\emph{\textbf{Lemma 4}} \cite{24}. For any $(K, F, Z, S)$ PDA $\textbf{P}$, if there exist $Z'$
useless $``\ast"$s in each column, we can obtain an $(F-Z')$-division $(K, M, N)$ coded caching scheme with a memory ratio of $$\frac{M}{N}=\frac{Z-Z'}{F-Z'}$$
and a transmission rate of $$R =\frac{S}{F-Z'},$$ in which the coding gain at each time slot is the same as the original scheme realized by
$\textbf{P}$ and Algorithm 1.

 Note that the operation field size $q$ of {\it Lemma 4} is $\mathcal{O}(F)$, implying that the size of each packet of files should approximate to $\log_2F$ bits. Therefore, the size of files in the server must be greater than $(F-Z')\log_2F$ so that the transmission rate of $\frac{S}{F-Z'}$ can be achieved. Given a $(K,F,Z,S)$ PDA, $\frac{Z}{F}>\frac{Z-Z'}{F-Z'}$ and $F>F-Z'$ always hold for any $Z,F,Z'\in\mathbb{N}^{+}$. Therefore, the scheme in {\it Lemma 4} has a smaller memory ratio and subpacketization level than that of {\it Lemma 1}. Furthermore, the scheme in {\it Theorem 5} can be improved as follows.



\emph{\textbf{Theorem 6}}. Given any $w,n_i\in \mathbb{N}^{+}$ and distinct positive integers $p_0, p_1, \ldots, p_{m-1}$ with $p_i\geq2$ and $w< n_0+n_1+\cdots+n_{m-1}$ for $i\in[0:m-1]$, there always exists a $[p_{0}^{n_0}
p_{1}^{n_1}\cdots p_{m-1}^{n_{m-1}}-(1+\sum_{i=1}^{w-1}\sum_{\substack{\mathcal{A}\subseteq\mathcal{X},|\mathcal{A}|=i}}\prod_{p^{(\beta)}_{\alpha}\in\mathcal{A}}
(p_{\alpha}-1))]$-division $(p_{0}^{n_0}p_{1}^{n_1}\cdots p_{m-1}^{n_{m-1}}, M, N)$ coded
caching scheme with a memory ratio of
$$\frac{M}{N}=1-\frac{\sum_{\substack{\mathcal{A}\subseteq\mathcal{X},|\mathcal{A}|=w}}\prod_{p^{(\beta)}_{\alpha}\in\mathcal{A}}(p_{\alpha}-1)}{p_{0}^{n_0}
p_{1}^{n_1}\cdots p_{m-1}^{n_{m-1}}-(1+\sum_{i=1}^{w-1}\sum_{\substack{\mathcal{A}\subseteq\mathcal{X},|\mathcal{A}|=i}}\prod_{p^{(\beta)}_{\alpha}\in\mathcal{A}}
(p_{\alpha}-1))},$$ and a transmission rate of $$R=\frac{\sum_{\textbf{e}\in\{\textbf{x}-\textbf{y}\mid (\textbf{x},\textbf{y})\in\mathcal{E}\}}p_{i_0}p_{i_1}\cdots p_{i_{w-1}}} {p_{0}^{n_0}p_{1}^{n_1}\cdots p_{m-1}^{n_{m-1}}-(1+\sum_{i=1}^{w-1}\sum_{\substack{\mathcal{A}\subseteq\mathcal{X},|\mathcal{A}|=i}}\prod_{p^{(\beta)}_{\alpha}\in\mathcal{A}}
(p_{\alpha}-1))},$$ where $$\mathcal{X}=\{\underbrace{p^{(0)}_0, p^{(1)}_0, \ldots, p^{(n_0-1)}_0}_{n_0}, \underbrace{p^{(0)}_1, p^{(1)}_1, \ldots, p^{(n_1-1)}_1}_{n_1}, \ldots, \underbrace{p^{(0)}_{m-1}, p^{(1)}_{m-1}, \ldots
, p^{(n_{m-1}-1)}_{m-1}}_{n_{m-1}}\},$$ and integers $i_0, i_1, \ldots, i_{w-1}$ are determined by $\mathcal{C}_\textbf{e}=\{(j_{\alpha_0}^{(i_0)}, n_{i_0}), (j_{\alpha_1}^{(i_1)} n_{i_1}),\ldots, (j_{\alpha_{w-1}}^{(i_{w-1})}, n_{i_{w-1}})\}$.
\begin{proof}
Let \textbf{P} denote an array generated from the injective arc-colored regular digraph $\hat{D}$. Based on the proof of {\it Proposition 1}, it can be seen that a $``\ast"$ in entry $\textbf{P}(\textbf{x}, \textbf{y})$ is useless if and only if $d_{\rm H}({\textbf{x}, \textbf{y}})<w$, where $\textbf{x}, \textbf{y}\in\mathcal{V}$. Therefore, the number of useless $``\ast"$s in each column of \textbf{P} is $$1+\sum_{i=1}^{w-1}\sum_{\substack{\mathcal{A}\subseteq\mathcal{X},|\mathcal{A}|=i}}\prod_{p^{(\beta)}_{\alpha}\in\mathcal{A}}(p_{\alpha}-1).$$ based on the results of {\it Lemma 4} and {\it Theorem 5}, the conclusion can be reached.
\end{proof}

Based on {\it Corollary 2} and {\it Lemma 4},
the following corollary can be obtained, which can be seen as a special case of {\it Theorem 6}. As its proof is similar to that
of {\it Theorem 6}, it is omitted.

\emph{\textbf{Corollary 4}}. Given any $n_0, w, p_0\in\mathbb{N}^{+}$ with $p_0\geq2$ and $w<n_0$, there always exists a
$[p_{0}^{n_0}-\sum_{i=0}^{w-1}{n_0\choose i}(p_0-1)^{i}]$-division $(p_{0}^{n_0}, M, N)$ coded caching scheme with a memory ratio of $$\frac{M}{N}=1-\frac{{n_0\choose w}(p_0-1)^{w}}{p_{0}^{n_0}-\sum_{i=0}^{w-1}{n_0\choose i}(p_0-1)^{i}},$$ and a transmission rate of $$R=\frac{{n_0\choose w}p_{0}^{w}(p_0-1)^{w}}{p_{0}^{n_0}-\sum_{i=0}^{w-1}{n_0\choose i}(p_0-1)^{i}}.$$

Finally, it should be pointed out that
if we consider a regular digraph with vertex set $\mathcal{V}=\mathbb{Z}_2^{n}$ $($or $\mathbb{Z}_3^{n})$ and arc set $\mathcal{E}=\{(\textbf{x},\textbf{y}), (\textbf{y},\textbf{x})\mid \textbf{x},\textbf{y}\in\mathcal{V}, d_{\rm H}(\textbf{x},\textbf{y})=w\}$,
the PDA constructions proposed in \cite{50} can also be viewed as an application of {\it Theorem 1}. The authors designed an appropriate partition for the entries of an array to satisfy the PDA constraints. Its partition rule can be seen as the equivalent class of edges defined in \cite{555}. Unlike previous existing constructions, our proposed construction of PDAs depends on the injective arc-colored regular digraphs, and the corresponding arc partition rule is different with the one of \cite{555}. Furthermore, our proposed coded caching scheme in {\it Theorem 6} extends the scheme of \cite{50} to the case with a flexible number of users.
\section{Performance Analyses of the New Schemes}
This section analyzes the proposed coded caching schemes in terms of the subpacketization level and transmission rate. They are compared with the existing schemes in Table I.
\subsection{Comparison between the Schemes in Theorem 4 and \cite{22}}
We compare our proposed scheme in {\it Theorem 4} with the scheme of \cite{22} in
Table II. Note that the
schemes in {\it Theorem 4} and \cite{22} are parameterized by $(n)$ and $(n,m,k,q)$, respectively. It can be seen that with the same number of users, subpacketization level, and memory ratio, our proposed scheme in {\it Theorem 4} has an advantage in the transmission rate. Moreover, for the same number of users and a slightly smaller memory ratio, our proposed scheme yields a smaller subpacketization level. But they are realized at the cost of some transmission rate.
\begin{table*}[!htbp]
\center
\caption{Comparison between the Scheme in Theorem 4 and the Scheme in \cite{22}}
\renewcommand\arraystretch{1}
  \setlength{\tabcolsep}{0.40mm}{
\begin{tabular}{|c|c|c|c|c|c|}
\hline
Schemes & Parameters & $K$ &$F$ & $\frac{M}{N}$ &  $R$ \\
\hline
$(n,m,k,q)$ in \cite{22}  & $(2,2,4,2)$ & $105$ & $105$ & $0.54$ & $8.0$\\
$(n)$ in {\it Theorem 4}  & $(105)$ & $105$ & $105$ & $0.54$ & $6.0$ \\
\hline
$(n,m,k,q)$ in \cite{22} & $(2,3,5,2)$ & $465$ & $4340$ & $0.59$ & $19.2$\\
$(n)$ in {\it Theorem 4}  & $(465)$ & $465$ & $465$ & $0.48$ & $30.0$ \\
\hline
$(n,m,k,q)$ in \cite{22} & $(2,4,6,2)$ & $1953$ & $546840$ & $0.61$ & $51.2$\\
$(n)$ in {\it Theorem 4}  & $(1953)$ & $1953$ & $1953$ & $0.45$ & $135.0$ \\
\hline
$(n,m,k,q)$ in \cite{22} & $(1,5,6,2)$ & $63$ & $5249660$ & $0.49$ & $5.3$\\
$(n)$ in {\it Theorem 4} & $(63)$ & $63$ & $63$ & $0.43$ & $9.0$ \\
\hline
\end{tabular}}
\end{table*}
\subsection{Comparison between the Schemes in Corollary 3, \cite{14} and \cite{80000}}

We discuss the performance of our scheme in {\it Corollary 3} by comparing it with the ones
of \cite{14} and \cite{80000}. The parameters of the schemes in {\it Corollary 3}, \cite{14} and \cite{80000} are
written as $(m,a,b, \lambda)$, $(n_0,w)$ and $(r, k, z)$, respectively. Table III shows that in comparison with the schemes of \cite{14}, our proposed scheme in {\it Corollary 3} yields a smaller subpacketization level, a slightly smaller memory ratio and a lower transmission rate for some parameters. Meanwhile, it is capable to serve more users simultaneously. When comparing with the scheme of \cite{80000},
with the same number of users and subpacketization level, our proposed scheme in {\it Corollary 3} has transmission rate advantage.
\begin{table*}[!htbp]
\center
\caption{Comparison between the Scheme in Corollary 3 and the Schemes in \cite{14,80000}}
\renewcommand\arraystretch{1}
  \setlength{\tabcolsep}{0.40mm}{
\begin{tabular}{|c|c|c|c|c|c|}
\hline
Schemes & Parameters & $K$ & $F$ & $\frac{M}{N}$ &  $R$ \\
\hline
$(m,a,b, \lambda)$ in \cite{14}  & $(16,12,10,6)$ & $1820$ & $8008$ & $0.88$ & $210.0$\\
$(r, k, z)$ in \cite{80000}  & $(5,256,1)$ & $4096$ & $4096$ & $0.81$ & $95.7$\\
$(n_0,w)$ in {\it Corollary 3}  & $(13,7)$ & $4096$ & $4096$ & $0.81$ & $23.2$ \\
\hline
$(m,a,b, \lambda)$ in \cite{14}  & $(20,5,6,3)$ & $15504$ & $38760$ & $0.88$ & $4.0$\\
$(r, k, z)$ in \cite{80000}  & $(6,512,1)$ & $32768$ & $32768$ & $0.89$ & $56.0$\\
$(n_0,w)$ in {\it Corollary 3}  & $(15,8)$ & $32768$ & $32768$ & $0.80$ & $25.1$ \\
\hline
$(m,a,b, \lambda)$ in \cite{14}  & $(20,14,12,7)$ & $38760$ & $125970$ & $0.84$ & $792.0$\\
$(r, k, z)$ in \cite{80000}  & $(6,1024,1)$ & $65536$ & $65536$ & $0.89$ & $111.9$\\
$(n_0,w)$ in {\it Corollary 3}  & $(16,9)$ & $65536$ & $65536$ & $0.83$ & $87.6$ \\
\hline
$(m,a,b, \lambda)$ in \cite{14}  & $(26,16,12,5)$ & $5311700$ & $9657700$ & $0.95$ & $5148.0$\\
$(r, k, z)$ in \cite{80000}  & $(7,65536,1)$ & $8388608$ & $8388608$ & $0.94$ & $4096.0$\\
$(n_0,w)$ in {\it Corollary 3}  & $(23,15)$ & $8388608$ & $8388608$ & $0.94$ & $1338.0$ \\
\hline
\end{tabular}}
\end{table*}

\subsection{Comparison between the Schemes in Corollary 4 and \cite{14,50}}
We first consider the comparison between the schemes in {\it Corollary 4} and \cite{50}.
With $p_0=2$, a coded caching scheme in {\it Corollary 4} can yield
$$K=2^{n_0},\; \frac{M}{N}=1-\frac{{n_0\choose w}}{\sum_{i=w}^{n_0}{n_0\choose i}},\;
 F=\sum_{i=w}^{n_0}{n_0\choose i},R=\frac{{n_0\choose w}2^{w}}{\sum_{i=w}^{n_0}{n_0\choose i}}.$$
By letting $w'=n_0-w$ for the scheme of \cite{50} (which is also shown in Table I), it can be observed that when $p_0=2$, it will be the same as
the scheme in {\it Corollary 4}. Therefore, our scheme proposed in {\it Corollary 4} generalizes the scheme
of \cite{50} and accommodates a more flexible number of users.

We further compare our proposed scheme in {\it Corollary 4} with the scheme of \cite{14} in
Table IV. The parameters of the
scheme in {\it Corollary 4} is parameterized by $(n_0,w,p_0)$. It can be seen that the proposed scheme in {\it Corollary 4} has a smaller
subpacketization level, a slightly smaller memory ratio and a lower transmission rate. Meanwhile, it can support more users.
\begin{table}[!htbp]
\center
\caption{Comparison between the scheme in Corollary 4 and the Scheme in \cite{14}}
\renewcommand\arraystretch{1}
  \setlength{\tabcolsep}{0.40mm}{
\begin{tabular}{|c|c|c|c|c|c|}
\hline
Schemes & Parameters & $K$ & $F$ & $\frac{M}{N}$ &  $R$ \\
\hline
$(n_0,w,p_0)$ in {\it Corollary 4} & $(7,6,6)$ & $279936$ & $187500$ & $0.42$ & $27216.0$\\
$(m,a,b, \lambda)$ in \cite{14}  & $(40,36,34,30)$ & $91390$ & $3838380$ & $0.49$ & $46376.0$ \\
\hline
$(n_0,w,p_0)$ in {\it Corollary 4} & $(8,7,6)$ & $1679616$ & $1015625$ & $0.38$ & $172270.0$\\
$(m,a,b, \lambda)$ in \cite{14}  & $(54,50,48,44)$ & $31625$ & $2587165$ & $0.38$ & $194580.0$ \\
\hline
$(n_0,w,p_0)$ in {\it Corollary 4} & $(8,6,4)$ & $65536$ & $44469$ & $0.54$ & $1880.1$\\
$(m,a,b, \lambda)$ in \cite{14}  & $(26,22,20,16)$ & $14950$ & $230230$ & $0.68$ & $4845.0$ \\
\hline
$(n_0,w,p_0)$ in {\it Corollary 4} & $(6,4,5)$ & $15625$ & $14080$ & $0.73$ & $170.5$\\
$(m,a,b, \lambda)$ in \cite{14} & $(20,16,14,10)$ & $4845$ & $38760$ & $0.79$ & $1001.0$ \\
\hline
\end{tabular}}
\end{table}
\section{Conclusion}
This paper has investigated the design of a PDA with the same number of rows and columns through the perspective of graph coloring, i.e., the injective arc coloring of a regular digraph. From this perspective, designing coded caching schemes with a linear subpacketization level can be converted into determining the number of colors that allow an injective arc coloring to the regular digraphs.
Based on this comprehension, we have defined a new class of regular digraphs and derived the upper bounds for the digraphs' injective chromatic index. Consequently, some new coded caching schemes that can support a flexible number of users have been obtained with a linear subpacketization level and a small transmission rate.


\bibliographystyle{ieeetr}

\end{document}